\documentclass[useAMS,usenatbib,epsfig, a4paper,amsmath,aps]{mnras}
 
\usepackage{float}
\usepackage{epsfig}
\usepackage{subfig}
\usepackage{epstopdf}
\usepackage{verbatim}
\usepackage{nicefrac}
\usepackage{url}
\usepackage{mathrsfs}
\usepackage{amssymb}
\usepackage{amsmath}

\newcommand{\kms}{km\,s$^{-1}$}

\title[Cosmology with Peculiar Velocities: Observational Effects]
   {Cosmology with Peculiar Velocities: Observational Effects}
\author[P.~Andersen et al.]
{P.~Andersen$^{1,2}$\thanks{email: \href{mailto:perandersen@dark-cosmology.dk}{\nolinkurl{perandersen@dark-cosmology.dk}}}, T. M. Davis$^{2,3}$,  C. Howlett$^{2,4}$ \\ 
$^{1}$Dark Cosmology Centre, University of Copenhagen, Copenhagen, Denmark. \\
$^{2}$ARC Centre of Excellence for All-sky Astrophysics (CAASTRO) \\
$^{3}$School of Mathematics \& Physics, The University of Queensland, St. Lucia, Brisbane, 4072, Australia.\\
$^{4}$International Centre for Radio Astronomy Research, The University of Western Australia, Crawley, WA 6009, Australia.
}
\date{\today}
\pubyear{2016}
\begin{document}
\maketitle

\begin{abstract}
In this paper we investigate how observational effects could possibly bias cosmological inferences from peculiar velocity measurements.  Specifically, we look at how bulk flow measurements are compared with theoretical predictions.  Usually bulk flow calculations try to approximate the flow that would occur in a sphere around the observer. Using the Horizon Run 2 simulation we show that the traditional methods for bulk flow estimation can overestimate the magnitude of the bulk flow for two reasons: when the survey geometry is not spherical (the data do not cover the whole sky), and when the observations undersample the velocity distributions. Our results may explain why several bulk flow measurements found bulk flow velocities that \textit{seem} larger than those expected in standard $\Lambda$CDM cosmologies.  We recommend a different approach when comparing bulk flows to cosmological models, in which the theoretical prediction for each bulk flow measurement is calculated specifically for the geometry and sampling rate of that survey.  This means that bulk flow values will not be comparable between surveys, but instead they are comparable with cosmological models, which is the more important measure.
\end{abstract}


\section{Introduction}
The term bulk flow in the context of cosmology refers to the average motion of matter in a particular region of space relative to the dipole subtracted cosmic microwave background (CMB) rest frame. One reason why bulk flows are interesting to cosmologists is that by measuring them we can learn more about the composition of the universe, the laws of gravity, and whether our current cosmological model is a good representation of the actual underlying dynamics.\\

A bulk flow is induced by density fluctuations, and thus the bulk motion we observe should match what we expect from the density distribution. The density distribution is in turn determined by cosmological parameters such as the strength of clustering, through $\sigma_8$, and the matter density, $\Omega_{\rm M}$. The magnitude of bulk flows can be predicted from theory given a model and set of cosmological parameters (e.g. $\sigma_8$ and $\Omega_{\rm M}$), some initial conditions (such as a fluctuation amplitude at the end of inflation), and a law of gravity (such as general relativity). If the observed bulk flow was to deviate from that predicted by theory, that would indicate that one or more of the given inputs is incorrect.\\

Currently tension exists in measurements of the bulk flow, with some measurements in apparent agreement with that predicted by $\Lambda$CDM \citep{2011MNRAS.414..264C, 2011JCAP...04..015D, 2011ApJ...736...93N, 2011ApJ...737...98O,  2012MNRAS.420..447T, 2013MNRAS.430.1617L, 2013MNRAS.428.2017M, 2014JCAP...09..019F, 2014MNRAS.437.1996M, 2014A&A...561A..97P, 2014MNRAS.445..402H, 2015MNRAS.450..317C} while others are not \citep{2008ApJ...686L..49K, 2009MNRAS.392..743W, 2010MNRAS.407.2328F, 2012MNRAS.419.3482A, 2015MNRAS.447..132W}. Relieving this tension is important if we are to gain physical insight into the nature of dark energy and dark matter.\\

The field of using large scale bulk flows to constrain cosmology has historically been limited by systematics due to the limited quality and quantity of the data available. Modern datasets now include peculiar velocity measurements of thousands of galaxies with moderate precision and hundreds of type Ia supernovae (SNe) with excellent precision. These have inspired a new generation of bulk flow studies. As these new datasets become increasingly abundant and precise, it is prudent to investigate the observational effects that may bias a bulk flow measured from one of these datasets.\\

One such effect is undersampling of the surveyed volume. Undersampling is especially relevant for estimates utilising a small number of distance indicators, like many recent estimates of the bulk flow done with observations of type Ia SNe \citep{2007ApJ...661..650H, 2007ApJ...659..122J, 2011MNRAS.414..264C, 2011JCAP...04..015D, 2011ApJ...732...65W, 2012MNRAS.420..447T, 2013A&A...560A..90F}. Attempts at addressing sampling issues have been proposed, see e.g. \cite{2009MNRAS.392..743W}, \cite{2012ApJ...761..151L} or \cite{2011ApJ...732...65W}. Another such effect is the geometry of a survey -- namely whether the survey covers the whole sky or a narrow cone.  Methods such as the minimum variance method proposed by \cite{2009MNRAS.392..743W} attempt to weight arbitrarily shaped survey geometries so that the bulk flow they calculate approximates what would have been measured if the distribution of data was spherical. Other effects, besides observational, might also play an important role. See e.g. \cite{2015JCAP...12..033H} where the effects of velocity correlations between supernova magnitudes are included in the data covariance matrix, and are found to have a significant impact on the constraints from a derived bulk flow estimate.\\

The bias that might arise from estimating the bulk flow magnitude with a small number of peculiar velocities, effectively undersampling the surveyed volume, and with a non-spherical distribution of measurements, is the focus of this paper. We utilise data from the Horizon Run 2 \citep[HR2;][]{2011JKAS...44..217K} simulation to investigate how strong a bias undersampling introduces for various survey volumes, from spherically symmetric surveys, to hemispherical and narrow cone surveys. We focus on the Maximum Likelihood (ML) estimator of the bulk flow, as it is computationally cheap to perform, easy to interpret and used widely in the literature. Additionally, for a limited test case, we investigate how successful the Minimum Variance (MV) \citep{2009MNRAS.392..743W} estimator is at alleviating the bias that comes from undersampling.  The ML and MV estimators are described in Appendix~\ref{app:mlemv}, where we take the opportunity to clarify some typographic errors and undefined terms in the original papers that can lead to confusion. \\

In section \ref{sec:hori2} we introduce the HR2 simulation. Then in section \ref{sec:lintheory} we summarise the theoretical footing of large scale bulk flows, and provide an expansion  beyond the usual spherical assumptions so that the theory is also valid for non-spherical geometries. The theoretical estimate is established as the benchmark against which we test the effects of undersampling. Then in section \ref{sec:samplingeffects} we analyse the effects of undersampling on the Maximum Likelihood estimator, for a spherical, hemispherical and narrow cone geometry. Finally in section \ref{sec:discussion} we discuss our findings and the implications for future work using large scale bulk flows in cosmology.\\

Throughout this paper when we refer to the theoretically most likely bulk flow magnitude it will be denoted the \emph{most probable} bulk flow magnitude, $V_p$, to avoid confusion with bulk flows from the Maximum Likelihood estimator.

\section{Simulation: Horizon Run 2}
\label{sec:hori2}
Throughout this paper we use the Horizon Run 2 (HR2) cosmological simulation \citep{2011JKAS...44..217K} to investigate how observational effects, in particular non-spherical survey geometries and undersampling, can influence bulk flow measurements in a $\Lambda$CDM universe. We choose this simulation for the following reason: the bulk motions of galaxies are primarily sensitive to large scale density perturbations, meaning that the bulk flow measured in apparently distinct patches drawn from a single simulation can remain significantly correlated. The HR2 simulation, containing 216 billion particles spanning a $(7.2h^{-1}\mathrm{Gpc})^3$ volume, is large enough that we can be confident our bulk flow measurements are effectively independent. The above simulation parameters result in a mass resolution of $1.25\times 10^{11}h^{-1}\mathrm{M}_{\odot}$, which allows us to recover galaxy-size halos with a mean particle separation of $1.2h^{-1}\,\mathrm{Mpc}$. The power spectrum, correlation function, mass function and basic halo properties match those predicted by WMAP5 $\Lambda$CDM \citep{2009ApJS..180..330K} and linear theory to percent level accuracy.\\

To generate our measurements we first draw spherical subsamples of radius $1h^{-1}\,\mathrm{Gpc}$ from the full HR2 dataset. The origin of each subset is chosen randomly, so that some will be chosen in higher than average density regions and some in lower than average density regions, incorporating the effects of cosmic variance. Knowledge of our local galactic surroundings could have been folded into the selection of origins, so that the subsets chosen would more closely represent the local environment that we find ourselves in. We have not done this, which means that the results of this work are the zero-knowledge results with no assumptions made about our position in the cosmological density field. 
In essence, we are comparing our one measurement of the bulk flow of our local universe to the distribution of bulk flows that $\Lambda$CDM would predict. It would also be enlightening to investigate whether there are any aspects of our local universe that would bias such a measurement, as \citet{2015JCAP...07..025W} did for supernova cosmology.  However, that is beyond the scope of this paper. \\

The HR2 subsets consist of approximately 3.1$\cdot 10^{6}$ dark matter haloes, each with six dimensional phase space information. Unfortunately a mock galaxy survey that fills the entire volume of the simulation does not exist, so in our analysis we assume that each DM halo corresponds to one galaxy. The smallest of the DM haloes are of a mass comparable to that of a galaxy, but the largest DM haloes of the HR2 simulation have a mass that would be equivalent to hundreds of galaxies. Effectively we are grouping galaxies in massive clusters into just one datapoint with the same probability of being subsampled as any other galaxy.\\

Fortunately, a limited number of mock SDSS-III \citep{2011AJ....142...72E} galaxy catalogues have been produced for the HR2 simulation, which allow us to test how this assumption may affect our results. In Appendix \ref{app:mockvsdm} we perform an analysis of the bulk flow magnitude distribution of galaxies from one such mock catalogue, and compare the distributions derived from the DM halo velocities. Our analysis shows that the distributions are similar, and, as such, treating each halo as an individual galaxy has minimal effect on our results.\\

To look at the effect of undersampling and non-spherical geometries, we wish to compare the actual bulk flow magnitude of a given number of galaxies within some volume, to the magnitude recovered using the ML and MV estimators. Although a real survey only has peculiar velocity information along the line-of-sight direction, both of these estimators attempt to reconstruct the 3D distribution of velocities and estimate the bulk flow. In this sense a fair comparison is then between the output of these estimators and the most probable bulk flow measured using the full 3D velocity vector for each galaxy. The method we use to determine the most probable bulk flow magnitude as well as the upper and lower 1-$\sigma$ limit for a particular subsample of the simulation is the following:
\begin{enumerate}
\item{Randomly place a geometry in the simulation.}
\item{Of total $N$ galaxies within the geometry, randomly draw $n$.}
\item{Derive the actual bulk flow vector of the $n$ galaxies, using the 3D velocity vector for each object.}
\item{Store the magnitude of the bulk flow vector.}
\item{Repeat the above process until the resulting distribution has converged.}
\end{enumerate}
Analogous to the method above we can determine the most probable bulk flow magnitude and 1-$\sigma$ upper and lower bounds for a specific bulk flow estimator, e.g. the ML estimator applied in section \ref{sec:samplingeffects}:
\begin{enumerate}
\item{Randomly place a geometry in the simulation.}
\item{Of total $N$ galaxies within the geometry, randomly draw $n$.}
\item{For the $n$ galaxies compute the line-of-sight velocities.}
\item{Apply the ML estimator to the line-of-sight velocities and derive the ML bulk flow vector.}
\item{Store the magnitude of the ML bulk flow vector.}
\item{Repeat the above process until the resulting distribution has converged.}
\end{enumerate}
The uncertainty associated with each peculiar velocity measurement is calculated as in Appendix A of \cite{2011ApJ...741...67D}, the implications of this are discussed in Appendix \ref{app:pecveluncer}. When determining upper and lower 1-$\sigma$ bounds we apply an equal likelihood algorithm, so that the 1-$\sigma$ limits are the equal likelihood bounds that encapsulate 68.27\% of the normalised distirbution.

\section{Linear Theory}
\label{sec:lintheory}
Under the assumption of the cosmological principle, that the universe is statistically isotropic and homogeneous, and assuming Gaussian density fluctuations, the velocity field at any given location can be treated as Gaussian random variate with zero mean and variance given by the velocity power spectrum $P_{vv}(k)$. Hence the bulk flow vector measured within some volume can also be described as a Gaussian random variate with zero mean and variance
\begin{equation}
\sigma^2_V(\boldsymbol{r}) = \int \frac{\mathrm{d}^{3}k}{(2 \pi)^3} P_{vv}(k) |\widetilde{W}(\boldsymbol{k};\boldsymbol{r})|^2.
\end{equation}
Assuming isotropy, this becomes
\begin{align}
\label{eq:rmsvar}
\sigma^2_V(\boldsymbol{r}) & = \frac{1}{2 \pi^2} \int_{k=0}^{\infty} \mathrm{d}k\,k^{2} P_{vv}(k) |\widetilde{W}(k;\boldsymbol{r})|^2, \notag \\
\implies \sigma^2_V(R) & = \frac{H_0^2 f^2}{2 \pi^2} \int_{k=0}^{\infty} \mathrm{d}k P(k) \widetilde{W}(k;R)^2,
\end{align}
where the Hubble constant, $H_0$, growth rate, $f$, and velocity and matter power spectra $P_{vv}(k)$ and $P(k)$ define a particular cosmology. The second equality of Eq.~\ref{eq:rmsvar}, which is commonly associated with the RMS velocity expected for a bulk flow vector \citep{2002coec.book.....C} follows from the assumption of a spherically symmetric window function and the linear approximation that $P_{vv} = H_0^2 f^2k^{-2}P_{\theta\theta}(k) \approx H_0^2 f^2k^{-2}P(k)$, where $P_{\theta\theta}$ is the power spectrum of the velocity divergence field (See Chapter 18 of \cite{2002coec.book.....C} for a review of the relationship between the density, velocity divergence and velocity fields, and \cite{2012MNRAS.427L..25J} for measurements of $P_{\theta\theta}$ from simulations). As can be seen in Fig. 1, $P_{\theta\theta}(k) = P(k)$ is typically a good assumption on the large scales probed by bulk flow measurements.\\
\begin{figure}
  \centering\includegraphics[width=\linewidth]{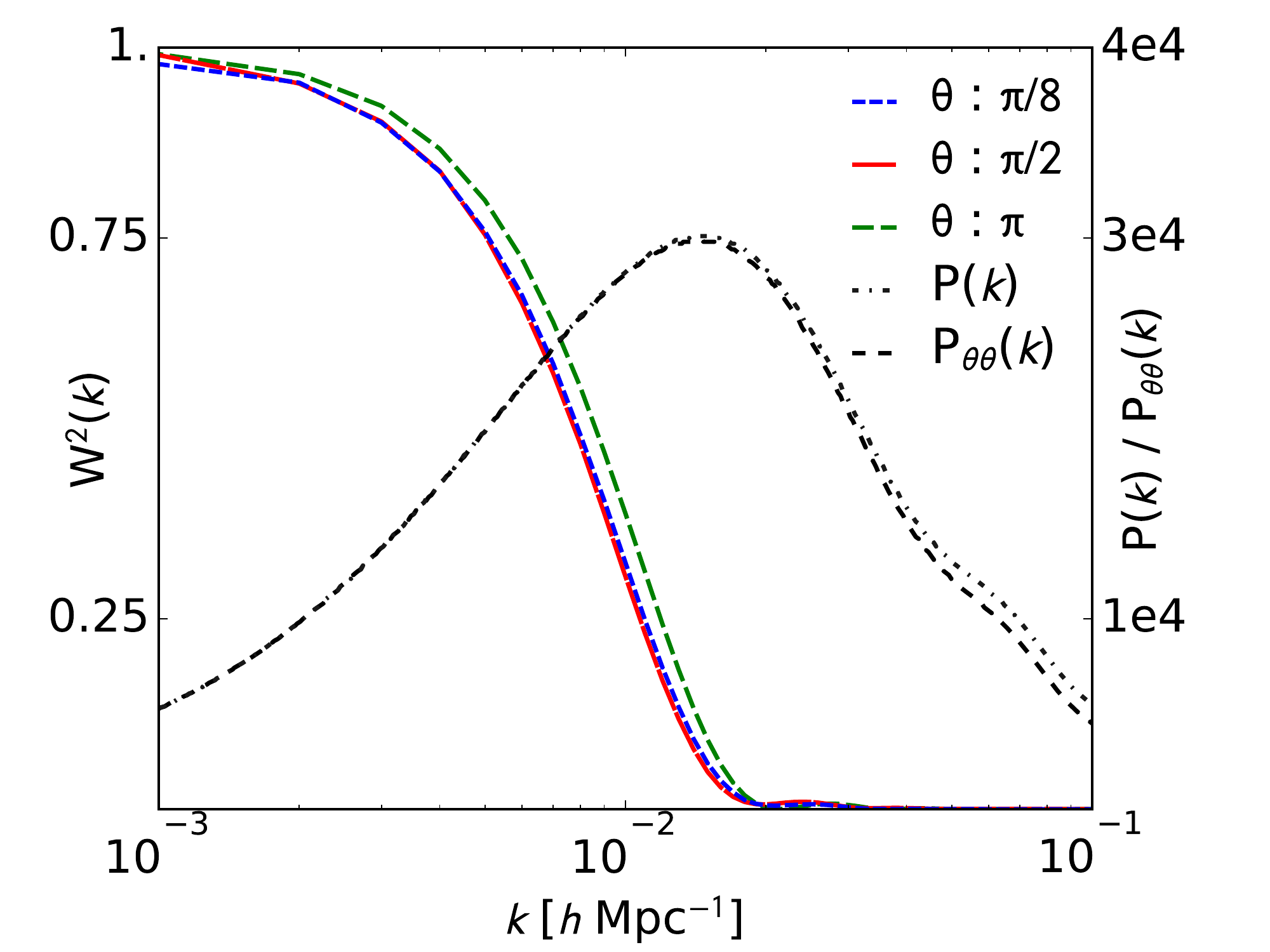}
  \caption{Window functions for the geometries used in this paper plotted along with the matter-matter and velocity divergence power spectra from {\sc copter}. $P(k)$ is the matter power spectrum, and $P_{\theta \theta}(k)$ the velocity divergence power spectrum. The geometries used are equal volume spherical cones with opening angles $\theta$, ranging from fully spherical $\theta=\pi$ to a very narrow cone with $\theta = \pi/8$}
  \label{fig:windowvspower}
\end{figure}

In Eq.~\ref{eq:rmsvar} $\widetilde{W}(k;\boldsymbol{r})$ is the Fourier transform of the window function, $W(\boldsymbol{r})$, for the geometry of the specific survey making that bulk flow measurement. The window function is a function of both $k$ and the volume in which the bulk flow is being measured. It measures how sensitive we are to measuring the statistical fluctuations at a particular scale. If the window function is large for a particular $k$ it means that we are highly sensitive to measuring fluctuations at the scale $k$ represents. The window function will be dependent on the geometry of the measurements taken to derive the bulk flow, and is therefore unique for each particular survey. For a fully spherical geometry of radius $R$ the window function takes the form
\begin{equation}
\label{eq:windowspherical}
\widetilde{W}(k;R) = \frac{3(\sin{kR} - kR\cos{kR})}{(kR)^3}.
\end{equation}
How strongly the window function of a particular survey will deviate from this spherical case will be determined by the geometry of the survey in question. Example window functions for conical geometries with a variety of opening angles are shown in Fig.~\ref{fig:windowvspower}. How these were calculated is detailed in section~\ref{sec:nonspherical}.\\

To calculate all the theoretical values of $\sigma_{V}$ in this paper we use a velocity divergence power spectrum, generated with the implementation of Renormalised Perturbation Theory \citep{2006PhRvD..73f3519C} in the {\sc copter} code \citep{2009PhRvD..80d3531C}. A linear {\sc camb}\footnote{http://camb.info/readme.html} \citep{2000ApJ...538..473L, 2012JCAP...04..027H} matter transfer function with the same cosmological parameters as HR2 and WMAP5 was used as input. From this {\sc copter} produces both a non-linear matter power spectrum as well as a non-linear velocity divergence power spectrum. We found that the difference between using the {\sc copter} velocity divergence power spectrum, the non-linear matter power spectrum, or the linear power spectrum was negligible, except for very narrow or small geometries where effects at $k \gtrsim 0.05$ become important. In Fig.~\ref{fig:windowvspower} we can see that for the geometries used in our analysis the differences when using the three power spectra are small as the spectra only differ in the regime where the window function vanishes. Nonetheless, throughout this paper we use the {\sc copter} velocity divergence power spectrum as that is most appropriate when working with bulk flows.\\

To calculate the theoretical most probable bulk flow magnitude $V_{p}(R)$ we use the fact that the peculiar velocity distribution is Maxwellian \citep{2012ApJ...761..151L} with RMS velocity $\sigma_V$, which gives us a probability distribution for the bulk flow amplitude of the form \citep{2002coec.book.....C}
\begin{equation}
p(V)\mathrm{d}V =  \sqrt{\frac{2}{\pi}}\left(\frac{3}{\sigma_V^2}\right)^{3/2} V^2 \exp{\left(-\frac{3V^2}{2\sigma_V^2}\right)} \mathrm{d}V.
\label{eq:Maxwellian}
\end{equation}
For this distribution the maximum probability value is then given by the relation
\begin{equation}
\label{eq:vbulkml}
V_{p}(R) = \sqrt{2/3} \, \sigma_V(R).
\end{equation}
When referring to the theoretical most probable bulk flow magnitude throughout this paper, it is this value based on a Maxwellian distribution of velocities that we are referencing. We confirmed that the velocities of halos in the HR2 simulation do indeed follow a Maxwellian distribution.\\

It is important to note that while the most probable bulk flow magnitude is a discrete value, it is still a value from a distribution with a variance. Optimally the theoretical distribution should be compared to an observed distribution of bulk flow magnitudes, but this is not practical in most situations. The best we can do is to compare our measured bulk flow magnitude with the most probable bulk flow magnitude from theory, but importantly remember to account for the variance on our theoretical prediction in our statistics.

\subsection{Non-Spherical Geometries}\label{sec:nonspherical}
As well as investigating the effects of undersampling on a spherical geometry, we wish to additionally develop a theoretical estimate for non-spherical geometries, that is we wish to break the assumption of spherical symmetry used to derive Eq.~\ref{eq:windowspherical}. For uniformly distributed surveys the window function takes the form
\begin{equation}
\label{eq:windowintegral}
\widetilde{W}(\boldsymbol{k};\mathbf{r}) = \frac{1}{V}\int_V \exp{(i\mathbf{k}\cdot \mathbf{r})}\mathrm{d}\mathbf{r},
\end{equation}
where $\exp{(i\mathbf{k}\cdot \mathbf{r})}$ can be expanded to \citep{2002coec.book.....C},
\begin{equation}
\exp{(i\mathbf{k}\cdot \mathbf{r})} = \sum_{l,m} j_l(kr)i^l(2l+1)\mathcal{P}_l^{|m|}(\cos{\theta})\exp{(im\phi)},
\end{equation}
where $\mathcal{P}_l^{|m|}$ are the Associated Legendre Polynomials.  The integral of Eq.~\ref{eq:windowintegral} then becomes
\begin{align}
\begin{split}
\int_V & \exp{(i\mathbf{k}\cdot \mathbf{r})}\mathrm{d}\mathbf{r} \\&= \sum_{l,m}i^l(2l+1) \int_0^{\phi_{max}} \exp{(im\phi)}\mathrm{d}\phi \\  &\int_0^{\theta_{max}}\mathcal{P}_l^{|m|}(\cos{\theta})\sin{\theta}\mathrm{d}\theta \int_0^R j_l(kr)r^2\mathrm{d}r
\label{eq:sphersum}
\end{split}
\end{align}
which for the spherical case where $(\theta_{max},\phi_{max})=(\pi,2\pi)$ reduces to Eq.~\ref{eq:windowspherical}. For a spherical cone geometry, with radius related to volume  and opening angle by
\begin{equation}
r = \left(\frac{3 V}{2 \pi (1-\cos{\theta})} \right)^{1/3},
\label{eq:spherconeradius}
\end{equation}
we can set $\phi_{max}=2\pi$ but let $\theta_{max}$ vary in the interval $(0;\pi]$. Regardless of the values of $l$, all terms of $m$ vanish except for the $m=0$ term. Therefore for non-spherical geometries we have to sum over $l$ to infinity. Although this approach is theoretically correct, in practice we would sum over $l$ only until the function value had converged to within computational accuracy. This is however very impractical since the complexity of the terms increase rapidly with $l$ making it difficult to include terms above $l\approx 20$. Unfortunately, we find that for our geometries that are very non-spherical only using terms $l\leq20$ is not sufficient to guarantee convergence. Hence this approach is still only practical for geometries close to a sphere.\\

Another approach to solving the window function for a given $k$ is to reformulate the volume integral in Cartesian coordinates
\begin{equation}
\widetilde{W}(k;\mathbf{r}) = \frac{1}{V}\int_0^{X}\int_0^{Y}\int_0^{Z}w(x,y,z)e^{i(kx+ky+kz)}\mathrm{d}x\mathrm{d}y\mathrm{d}z.
\label{eq:windowcartesian}
\end{equation}
The triple integral is over a cube that is at least large enough to contain the volume $V$ from Eq.~\ref{eq:windowintegral}. The $w(x,y,z)$ function is introduced, defined as being one inside the volume and zero otherwise, which makes sure the volume integrated over is conserved. The conversion to Cartesian coordinates makes it simpler to solve the integral numerically. It should be noted that even though we only consider rotationally symmetric windows with constant number density in this study, the above equation can be extended to include surveys of arbitrary geometry and non-constant number density, simply by choosing a suitable function $w(x,y,z)$.\\

Based on Eq.~\ref{eq:windowcartesian} we developed two pieces of code to solve the problem numerically, one calculating the integral using MCMC methods and the other applying a trapezoidal volume integral.\footnote{For details and link to the source code see https://github.com/per-andersen/MV-MLE-BulkFlow} The independence of the two codes is used to confirm the validity of the results; the outputs from the two codes are consistently within 3\% percent of one another.\\

To see how this theoretical prediction compares with the actual underlying bulk flow of the HR2 simulation, we plot the most probable bulk flow magnitude as well as the upper and lower 1-$\sigma$ limits as a function of geometry in Fig.~\ref{fig:bulkangle}. The geometry in this case is a spherical cone where the opening angle $\theta$ is varied. It is worth noting that the volume of the geometry is kept constant as the opening angle $\theta$ is varied. This is achieved by varying the radial extent of the geometry along with $\theta$ according to Eq.~\ref{eq:spherconeradius}. Keeping the volume constant helps keep the simulation and theoretical results almost constant as $\theta$ is varied. For all opening angles we see that our theoretical value matches that measured from the simulations extremely well.\\

\section{Geometry and Sampling Effects}
\label{sec:samplingeffects}
In this section we present how non-spherical geometries and undersampling of the cosmological volume can impact the results of the ML and MV estimators. We use the theoretically predicted most probable bulk flow magnitude as a benchmark; the closer the estimator comes to replicating the theoretical distribution the better.\\

\begin{figure}
  \centering\includegraphics[width=\linewidth]{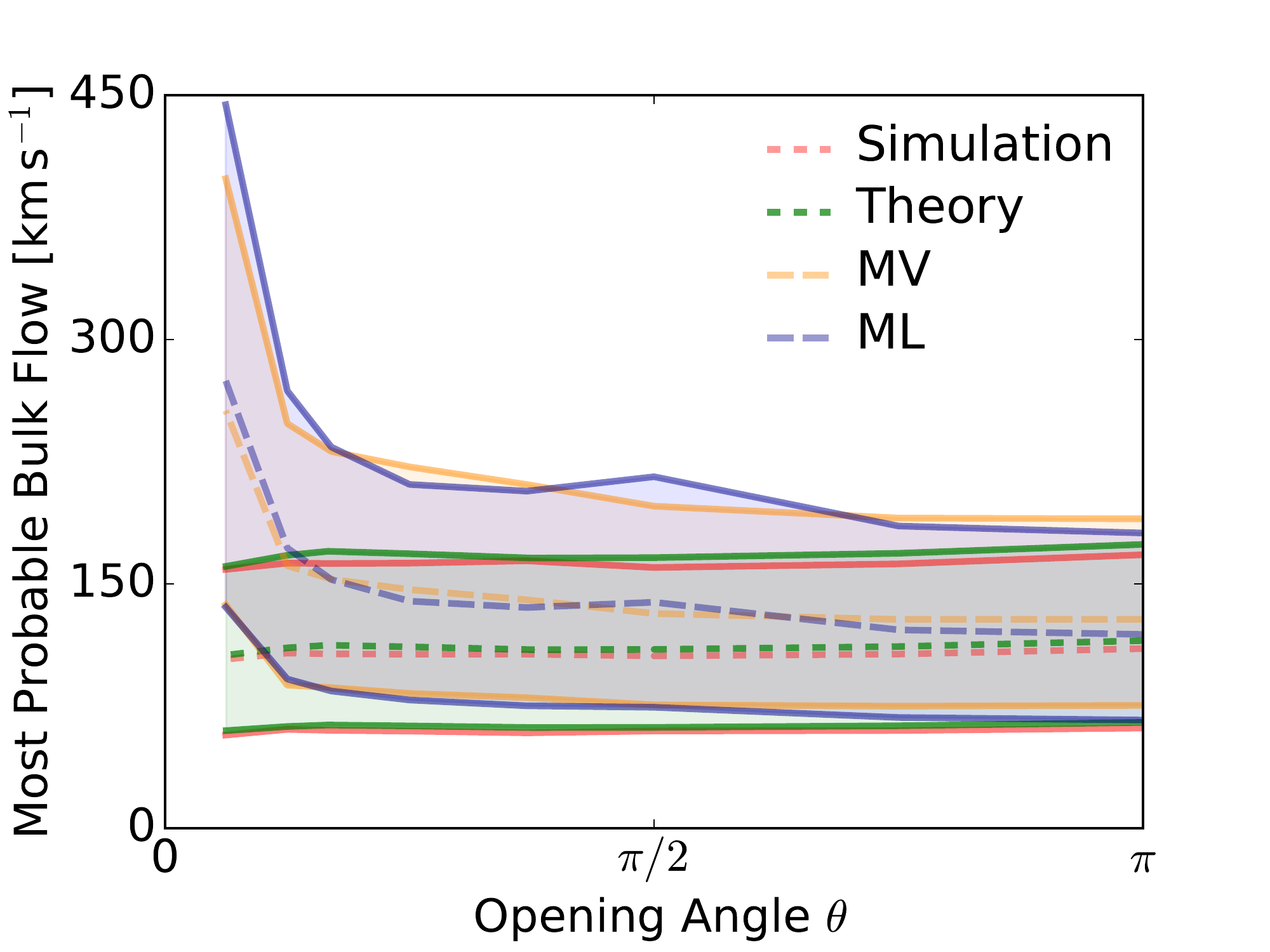}
  \caption{The most probable measured bulk flow magnitude as a function of opening angle for a spherical cone geometry. The tested geometries vary in opening angle from the fully spherical situation where $\theta=\pi$, over a hemisphere to the most narrow geometry tested being $\theta = \pi/16$. The distributions from the simulation, theory, MV estimator, and ML estimator are shown with the dashed line being the most probable bulk flow magnitude and the colored band showing the upper and lower 1-$\sigma$ limits. For both the ML and MV estimators the sampling was fixed at $n=500$. For the MV estimator the ideal radius $R_I$ was set to 50 Mpc $h^{-1}$.}
  \label{fig:bulkangle}
\end{figure}
We first investigate the scenario where we use a fixed number of objects ($n=500$) and compare both the performance of the ML and MV estimator. The results can be seen in Fig.~\ref{fig:bulkangle}. Both the ML and MV estimator have a bias towards measuring larger bulk flow magnitudes on average than the actual underlying bulk flows. As the survey geometry becomes narrower, however, this bias increases, with the most narrow geometry having the strongest bias. The behaviour of the ML and MV estimators is very similar.\\

In the narrow cone regime, both the ML and MV estimators predict significantly larger most probable bulk flows than would be expected from theory. Hence, incorrectly accounting for non-spherical geometries in the ML and MV estimators could potentially lead one to conclude they had measured a larger bulk flow than would be expected in a $\Lambda$CDM universe.\\

\begin{figure}
\centering\includegraphics[width=\linewidth]{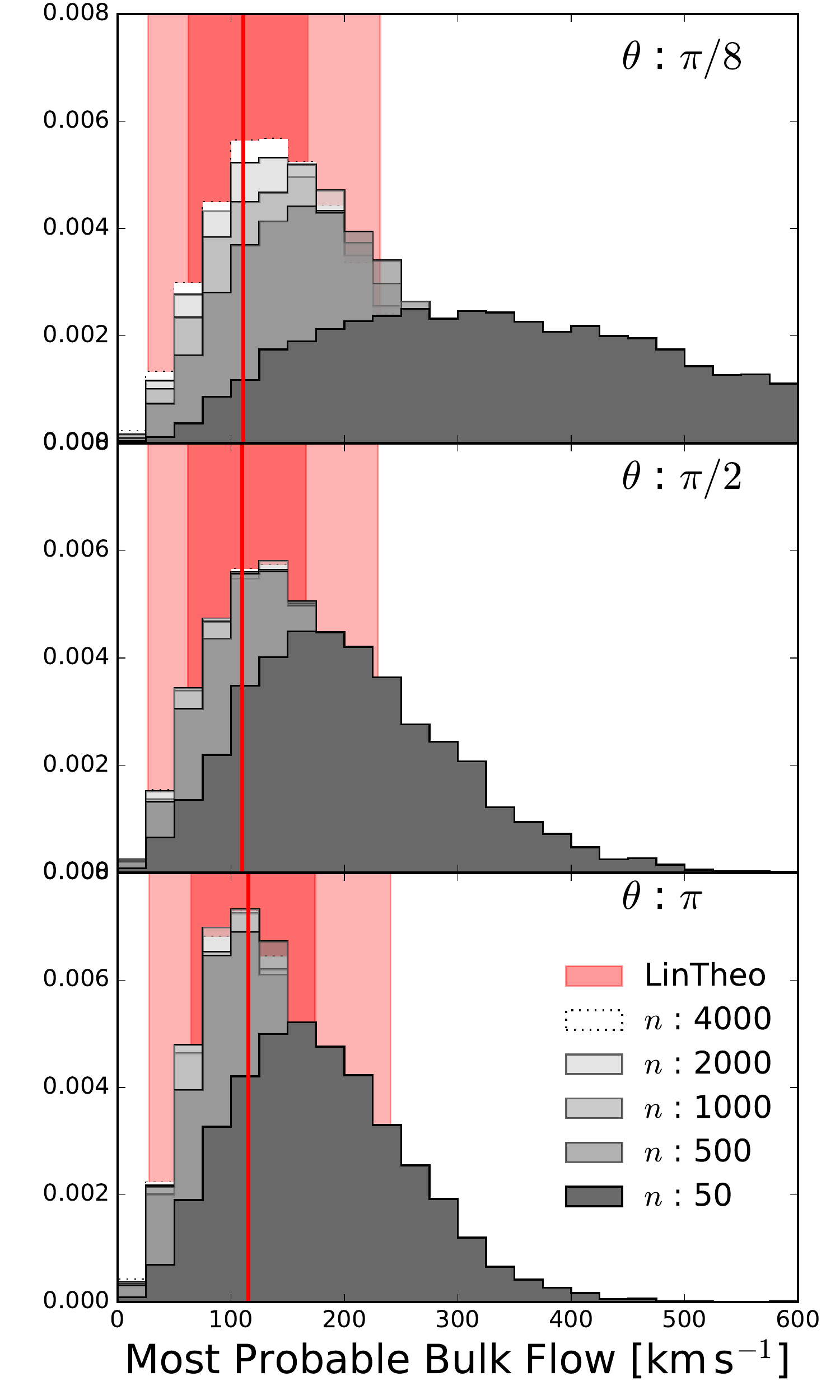}
\caption{Distributions of ML bulk flow magnitudes for various sampling rates, $n$. The top/middle/bottom distributions correspond to a geometry with opening angle $\frac{\pi}{8}$/$\frac{\pi}{2}$/$\pi$. The volume is kept constant as opening angle is varied, resulting in a radius of 631/267/210 Mpc~$h^{-1}$. }
\label{fig:cosmicsamplingcombined}
\end{figure}
Next we investigate how the sampling rate can create biases in the most probable bulk flow calculated using the ML and MV estimates for a fixed geometry. For the values $n \in [50, 500, 1000, 2000, 4000]$ and opening angles $\theta \in [\pi/8, \pi/2, \pi]$, corresponding to a narrow spherical cone, a hemisphere, and a full sphere, we apply the ML estimator as described in section \ref{sec:hori2}. The results can be seen in Fig.~\ref{fig:cosmicsamplingcombined}. There are two noteworthy trends from this plot. The first is that for all geometries the estimated most probable value is shifted to be 1-$\sigma$ away from the actual most probable value when the sampling is less than $n \lesssim 500$. The second is that this effect is stronger for narrow geometries, in our case the geometry with opening angle $\theta = \pi/8$ is much more adversely affected by undersampling than the hemispherical or spherical case. What this means in practice is that estimates of the bulk flow magnitude that utilise a small number of peculiar velocities are likely to be biased by undersampling effects in such a way that we would measure on average a larger bulk flow magnitude than the actual underlying bulk flow being probed. Of particular interest is the fact that this remains true even for spherical geometries if the number of objects is small.\\

The most probable bulk flow velocities for the distributions in Fig.~\ref{fig:cosmicsamplingcombined}, as well as a few additional configurations of sampling rate and opening angles, are listed in Table~\ref{tab:samplinggeometryeffectscombined}. The absolute differences between the most probable bulk flow values derived from simulation and theory are also listed. This absolute difference is an indicator of how strong a bias we might expect in the distribution of bulk flows derived for a particular sampling rate and survey geometry. A small absolute difference between most probable bulk flow velocities from simulation and theory indicates that the sampling rate is sufficient, and that minimal bias is to be expected for that particular survey geometry. In using Table~\ref{tab:samplinggeometryeffectscombined} it is important to note that not only the most probable bulk flow velocity, $V_p$, is shifted towards larger values. Rather, the entire distribution of bulk flow velocities is shifted, including the one and two sigma limits. Looking at, e.g., line seven of Table~\ref{tab:samplinggeometryeffectscombined} where $n=50$ and $\theta=0.125 \,\pi$ we see that even though the theory predicts something close to $\sim$100 km s$^{-1}$ a measured bulk flow value of $\sim$500 km s$^{-1}$ is still within the one sigma confidence limits, and hence is still well within the expectations of a $\Lambda$CDM cosmology.\\

\begin{figure}
\centering\includegraphics[width=\linewidth]{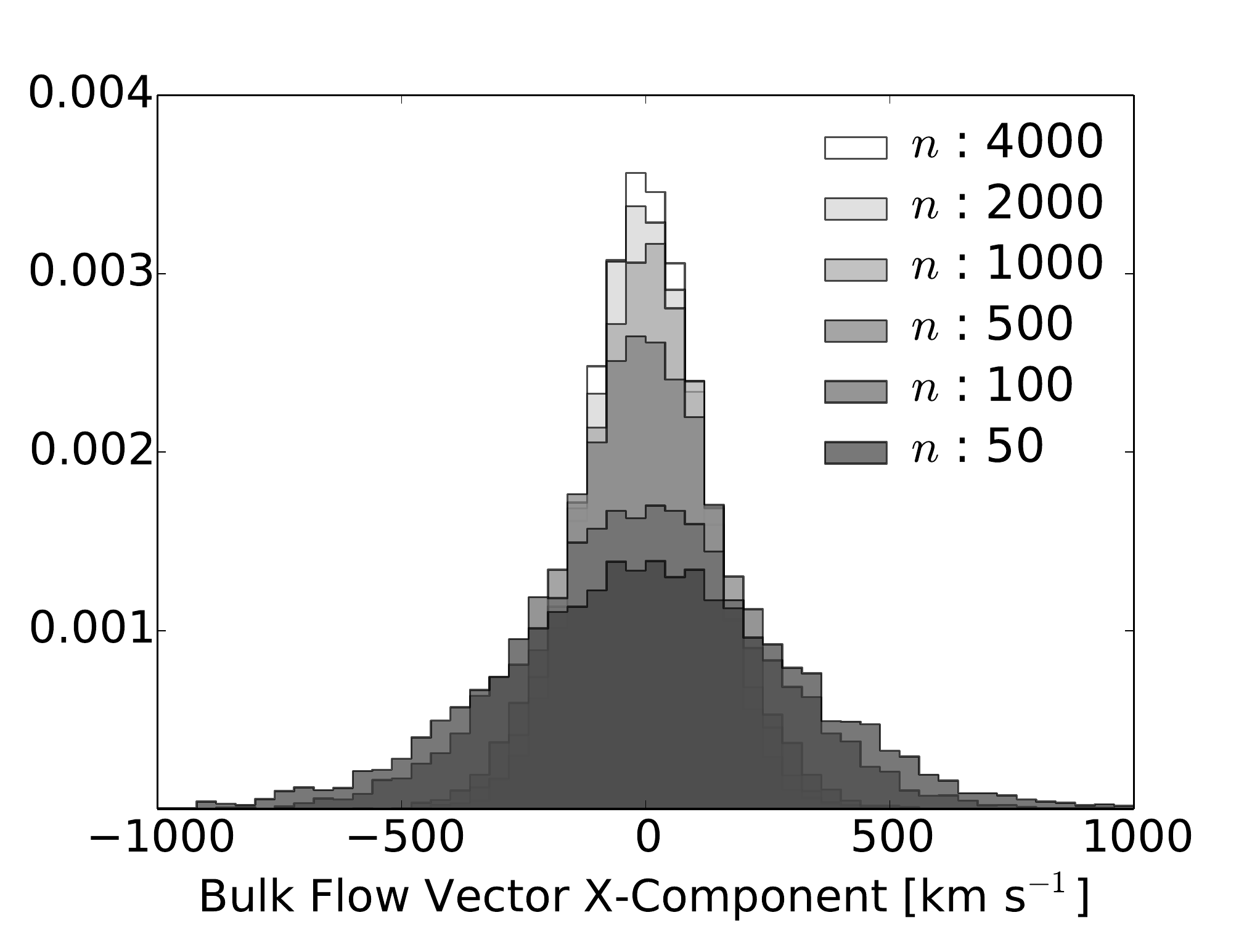}
\caption{The distribution of the $x$-components of the bulk flows from the top panel of Fig.~\ref{fig:cosmicsamplingcombined} where the opening angle is $\theta=\pi/8$. Poorer sampling leads to a larger variance in the Gaussian-distributed velocity components, which in turn causes the most probable bulk flow to shift to a larger value.}
\label{fig:cosmicvelocomponent}
\end{figure}

The cause for the bias from poor sampling is the increased variance of the bulk flow velocity components; in Fig.~\ref{fig:cosmicvelocomponent} the $x$-components of the bulk flow velocities from the top panel of Fig.~\ref{fig:cosmicsamplingcombined} are plotted for the various sampling rates. When sampling decreases variance increases, which in turn causes the most probable bulk flow value to shift according to Eq.~\ref{eq:vbulkml}. Note that $\sigma_{V}$ in Eq.~\ref{eq:vbulkml} denotes the variance of the bulk flow \textit{vector}, which is equal to the RMS because the distribution of bulk flow vectors is Gaussian.  The variance in any one Cartesian component of the bulk flow vector is then $\sigma_{V}/\sqrt{3}$.  For the bulk flow {\em magnitude} $\sigma_{V}$ refers only to the RMS, due to the relationship between Maxwellian and Gaussian distributions.  The variance of the bulk flow magnitude is then given by\footnote{To derive this use $p(V)dV$ from equation~\ref{eq:Maxwellian} in the definition of variance $\sigma^2\equiv\int_0^\infty p(V)(v-\bar{v})^2dV$, where the standard integral  $\int_0^{\infty}x^ne^{-bx^2}dx = \frac{(2k-1)!!}{2^{k+1}b^k}\sqrt{\frac{\pi}{b}}$ with $n=2k$ and $b>0$ comes in handy; note $(2k-1)!! \equiv \Pi_{i=1}^k(2i-1) = \frac{(2k)!}{2^kk!}$.  } $\sigma^2 = \sigma_{\rm V}^2(1-\frac{8}{3\pi})$.\\

Another way to illustrate this effect is to imagine a large volume where the galaxies within obey the cosmological principle such that if you sum over the velocities of all $N$ galaxies you will derive a bulk flow magnitude of exactly zero. This will be true even if only the line-of-sight components of the peculiar velocities are observed. If then only $n < N$ peculiar velocities are observed, it is very likely that a non-zero bulk flow magnitude will be measured, and since a magnitude can only ever be positive we are now dealing with some non-zero positive number. We might redraw a new set of $n$ galaxies and derive a different magnitude, but it is still going to be some non-zero positive number. If $n \approx N$ then we are likely to measure a magnitude that is closer to zero than if we only draw $n \ll N$ galaxies. In other words, undersampling \textit{always} increases our RMS velocity and skews the most probable measured magnitude towards larger values.

\begin{table*} 
 \centering 
 \begin{tabular}{| c | c | c | c | c | c |} 
 \hline $ $ & $\theta$ & V$_p$ & 68\% Limits&$\mid$V$_p$ - V$_{p,\mathrm{theory}}\mid$ & Sample Density\\
 $ $ & $ $ & km s$^{-1}$ & km s$^{-1}$ & km s$^{-1}$ & $(h^{-1}\mathrm{Mpc})^{-3}$\\ \hline
n : 8000 & 0.125 $ \, \pi$ & 132 $^{+73}_{-61}$ & 71 - 205 & 26 & 200$ \times 10^{-6}$ \\ 
n : 4000 & 0.125 $ \, \pi$ & 136 $^{+75}_{-63}$ & 73 - 210 & 29 & 100$ \times 10^{-6}$ \\ 
n : 2000 & 0.125 $ \, \pi$ & 142 $^{+79}_{-66}$ & 75 - 221 & 35 & 50$ \times 10^{-6}$ \\ 
n : 1000 & 0.125 $ \, \pi$ & 153 $^{+86}_{-72}$ & 80 - 238 & 46 & 25$ \times 10^{-6}$ \\ 
n : 500 & 0.125 $ \, \pi$ & 173 $^{+97}_{-81}$ & 92 - 269 & 66 & 12$ \times 10^{-6}$ \\ 
n : 100 & 0.125 $ \, \pi$ & 257 $^{+143}_{-119}$ & 137 - 399 & 150 & 2$ \times 10^{-6}$ \\ 
n : 50 & 0.125 $ \, \pi$ & 326 $^{+203}_{-166}$ & 160 - 528 & 219 & 1$ \times 10^{-6}$ \\ \hline \hline
n : 8000 & 0.5 $ \, \pi$ & 131 $^{+74}_{-62}$ & 68 - 205 & 21 & 200$ \times 10^{-6}$ \\ 
n : 4000 & 0.5 $ \, \pi$ & 131 $^{+74}_{-61}$ & 69 - 204 & 22 & 100$ \times 10^{-6}$ \\ 
n : 2000 & 0.5 $ \, \pi$ & 132 $^{+75}_{-63}$ & 69 - 207 & 22 & 50$ \times 10^{-6}$ \\ 
n : 1000 & 0.5 $ \, \pi$ & 133 $^{+76}_{-63}$ & 69 - 208 & 23 & 25$ \times 10^{-6}$ \\ 
n : 500 & 0.5 $ \, \pi$ & 136 $^{+77}_{-64}$ & 72 - 213 & 27 & 12$ \times 10^{-6}$ \\ 
n : 100 & 0.5 $ \, \pi$ & 158 $^{+89}_{-74}$ & 84 - 247 & 49 & 2$ \times 10^{-6}$ \\ 
n : 50 & 0.5 $ \, \pi$ & 180 $^{+96}_{-81}$ & 99 - 275 & 70 & 1$ \times 10^{-6}$ \\ \hline \hline
n : 8000 & 1.0 $ \, \pi$ & 110 $^{+58}_{-49}$ & 61 - 168 & 3 & 200$ \times 10^{-6}$ \\ 
n : 4000 & 1.0 $ \, \pi$ & 110 $^{+58}_{-49}$ & 61 - 167 & 3 & 100$ \times 10^{-6}$ \\ 
n : 2000 & 1.0 $ \, \pi$ & 111 $^{+59}_{-49}$ & 61 - 169 & 2 & 50$ \times 10^{-6}$ \\ 
n : 1000 & 1.0 $ \, \pi$ & 113 $^{+59}_{-50}$ & 62 - 171 & 0 & 25$ \times 10^{-6}$ \\ 
n : 500 & 1.0 $ \, \pi$ & 116 $^{+61}_{-51}$ & 64 - 176 & 3 & 12$ \times 10^{-6}$ \\ 
n : 100 & 1.0 $ \, \pi$ & 138 $^{+72}_{-61}$ & 77 - 210 & 25 & 2$ \times 10^{-6}$ \\ 
n : 50 & 1.0 $ \, \pi$ & 159 $^{+83}_{-70}$ & 89 - 241 & 46 & 1$ \times 10^{-6}$ \\  \hline
 \end{tabular}  
\caption{$V_p$ is the most probable bulk flow for the distribution of bulk flows derived from simulation using the ML estimator, for the given survey geometry, defined by the opening angle $\theta$, and sampling rate, given by $n$, which is the number of peculiar velocities per derived bulk flow estimate. The upper and lower one sigma equal likelihood limits encapsulating 68\% of the likelihood are also listed. $\mid$V$_p$ - V$_{p,\mathrm{theory}}\mid$ is the absolute difference between the most probable bulk flow velocity derived from estimate and from linear theory. A small absolute difference indicates that the sampling rate is sufficient for the given geometry, such that the derived distribution matches the actual underlying distribution. In the final column the survey sample density is listed for reference.} 
\label{tab:samplinggeometryeffectscombined} 
\end{table*}

\section{Discussion \& Conclusion}
\label{sec:discussion}
After reviewing linear theory we showed how it can be expanded to be valid for non-spherical geometries, developing code that numerically calculates the theoretical bulk flow magnitude for any arbitrary survey geometry. To test the validity of the developed code, the derived theoretical bulk flow magnitude was compared to that of a variety of spherical cone geometries in the Horizon Run 2 (HR2) cosmological simulation and found to be within 5\% or better agreement for all tested geometries.\\

However, when simulating more realistic surveys and applying the Maximum Likelihood (ML) estimator we found that undersampling effects severely bias measurements of the bulk flow magnitude when a small number ($n \lesssim 500$) of peculiar velocities are used in the bulk flow estimate. On average, undersampling pushes the measured bulk flow to higher values, with the bias being amplified when narrower survey geometries are used.\\

For our fixed volume of 40$\cdot 10^6 (h^{-1}\mathrm{Mpc})^3$ using 500 SNe corresponds to a sampling density of $\sim13 \, \mathrm{SNe}/10^6 (h^{-1}\mathrm{Mpc})^3$. Hence we expect undersampling could affect many recent measurements of the of the bulk flow magnitude utilising type Ia SNe as a distance indicator (i.e, \citealt{2007ApJ...661..650H, 2007ApJ...659..122J, 2011MNRAS.414..264C, 2011JCAP...04..015D, 2011ApJ...732...65W, 2012MNRAS.420..447T, 2013A&A...560A..90F}) where the number of supernovae are well below 300 and the sampling density is also well below $13\; \mathrm{SNe}/10^6 (h^{-1}\mathrm{Mpc})^3$.\\ 

Without a detailed analysis of each of the previous bulk flow estimates, which is beyond the scope of this paper, it is hard to determine whether or not a particular result is affected by undersampling. However, some examples that might deserve attention include  
e.g. \cite{2013A&A...560A..90F} where the SNe are subdivided into four shells, and for the SNe in each shell a bulk flow is estimated. 
We would expect the bulk flow to converge to the CMB frame as we go to higher redshifts and larger volumes, and yet \cite{2013A&A...560A..90F} find that in shells of both increasing redshift and increasing volume there is no clear trend in the magnitude of the bulk flow.  Instead, the trend they see could  potentially be explained by undersampling.  Their bins contain varying numbers of supernovae, namely $n=[128, 36, 38, 77]$, in which they find bulk flows of $V_{\rm p}=[243, 452, 650, 105]$\kms.  So there is a trend by which the bins with fewer supernovae find larger bulk flows (e.g. compare the middle two bins with the outer two bins).\\  

Similarly, \citet{2012MNRAS.420..447T} provide two measurements of the ML bulk flow: one with all 245 SNe from the First Amendment compilation, the other with a subset of 136 SNe that excludes the nearby ones (excludes $z<0.02$).  Naive expectations would suggest that the sample focussing on higher redshift SNe should be closer to converging on the CMB and thus have a lower bulk flow, however they find the opposite.  The higher-redshift-only sample has a higher bulk flow, but since it has fewer data points than the full sample that would  be consistent with our finding that undersampling overestimates the bulk flow. \\ 

Both \cite{2013A&A...560A..90F} and \citet{2012MNRAS.420..447T} found bulk flows that exceeded the predicted flow based on known density distributions in the nearby universe, so whether the estimates are inflated by undersampling is potentially an interesting question (although neither claimed significant deviation from $\Lambda$CDM).  While we have selected these two as the most significant examples that could be affected by the sampling biases we discuss in this paper, we note that this trend is pervasive, as no other samples show significant opposing trends. Some show slight reduction in bulk flow with  smaller samples, but it is much less significant than the positively correlated examples above (and much smaller than the uncertainties), e.g. in \cite{2011MNRAS.414..264C} increasing the sample from 61 to 109 SNe increases the estimated bulk flow from 250 \kms to 260 \kms, an effect of less than 5\%.\\   

For bulk flow estimates where the typical number of observed peculiar velocities in a survey is $n \gtrsim 3000$, i.e., most estimates using the Tully-Fisher or Fundamental Plane relation \citep{2011ApJ...736...93N,2014MNRAS.437.1996M, 2015MNRAS.447..132W, 2016MNRAS.455..386S}, we found no bias from undersampling. It is however important to note that the analysis of this paper assumes type Ia SNe are used as distance indicators, and therefore the uncertainties in each distance measurement are small (Appendix~\ref{app:pecveluncer}). The typically larger uncertainties derived from Tully-Fisher or Fundamental Plane estimates would increase the variance in the individual bulk flow components, which in turn could mean we require larger numbers of objects to avoid biases than is found here.\\

Effects from uneven sampling have previously been discussed in the literature. One example is Eq.~10 of \cite{2012ApJ...761..151L} where a method of dividing the measured peculiar velocities by their selection function is proposed. In \cite{1982ApJ...258...64A} and \cite{2007ApJ...661..650H} Monte Carlo simulations of observations are used to better understand systematic effects, including sampling effects. Other works \citep{2011ApJ...732...65W, 2009MNRAS.392..743W} develop new estimators such as the Weighted Least Squares (WLS), the Coefficient Unbiased (CU), or the Minimum Variance (MV) estimators, with the MV estimator being the most popular alternative to the ML estimator. The MV estimator is constructed in part to account for sampling bias (with the motivation to be able to compare measurements of bulk flow between surveys); in our work we found that the MV estimator suffered the same bias as the ML estimator, again with the bias increasing for narrower geometries.\\

A number of recent papers compare a measured bulk flow directly to a $\Lambda$CDM prediction based on linear theory and an assumption of spherical symmetry. For example \cite{2011MNRAS.414..264C}, \cite{2011JCAP...04..015D}, and \cite{2016MNRAS.455..386S} plot bulk flow measurements as a function of redshift compared to a generic $\Lambda$CDM prediction.  Our analysis suggests that such a comparison between bulk flows derived from different surveys, and therefore different survey geometries and sampling rates, is potentially problematic.\\

In \cite{2012ApJ...759L...7P} the HR2 simulation was used to show that the size of the large scale structure known as the Sloan Great Wall (SGW) is in agreement with what we statistically expect from $\Lambda$CDM cosmology, something that had previously been disputed. Similarly, as early as \cite{1982ApJ...258...64A} simulations were being used to compare measured bulk flows to theoretical predictions.  Analogous to their arguments, our study highlights the importance of considering the full distribution of bulk flow magnitudes from theory, including sampling effects, rather than focusing on only the most probable bulk flow magnitude.  
That is, we propose that bulk flows should not be compared to the prediction from linear theory, but with the bulk flow magnitude distribution derived from a cosmological simulation using the method described above, with the actual survey geometry given as input.

\section*{Acknowledgements}
Parts of this research were conducted by the Australian Research Council Centre of Excellence for All-sky Astrophysics (CAASTRO), through project number CE110001020. The Dark Cosmology Centre was funded by the DNRF.
\bibliographystyle{mnras}
\bibliography{BulkFlow}
\appendix
\section{ML and MV Bulk Flow Estimators}
\label{app:mlemv}
To compare the measured bulk flow with theoretical predictions, it is necessary to have a method to turn the individually observed peculiar velocities into a bulk flow. In this paper we focus on two estimators, the Maximum Likelihood (ML) and the Minimum Variance (MV) estimators. In the original paper introducing the MV estimator \citep{2009MNRAS.392..743W} there were a few typographic errors and unexplained terms; for completeness and to help others avoid confusion the procedures used to carry out the ML and MV estimators in this work are explained in this appendix. 
\subsection{Maximum Likelihood}
The ML estimator is by far the easiest of the two to implement and is computationally much cheaper than the MV estimator. The result of the ML estimator is a vector containing the velocity components corresponding to each of the three spatial dimensions. Each of the three components is given by a sum over the individual peculiar velocity components multiplied by some weight. The sum has the form
\begin{equation}
u_i = \sum_{n} w_{i,n} S_n
\end{equation}
where $i$ is the placeholder for either the $x$, $y$, or $z$ index and the sum goes over all $n$ peculiar velocities. $S_n$ is the $n$'th measured peculiar velocity, $w_{i,n}$ is the associated weight for that peculiar velocity and $u_i$ is the calculated bulk flow where again $i=(x,y,z)$. This equation holds true for both the ML and the MV estimators. Where they differ is how they go about calculating the $w_{i,n}$ weights.\\

For the ML estimate the weights are given by
\begin{equation}
w_{i,n} = \sum_j \frac{ \hat{x}_j \cdot \hat{r}_n}{(\sigma_n^2 + \sigma_\star^2)}A_{ij}^{-1}.
\end{equation}
The sum is over the $j=(x,y,z)$ components, and $\hat{x}_j \cdot \hat{r}_n$ is the projection of the unit vector $\hat{r}$ pointing from the observer to the galaxy in question. $\sigma_n$ is the uncertainty on the velocity of the $n$'th measurement, and $\sigma_\star$ is a constant of order 250 km s$^{-1}$ meant to account for the non-linear flows on smaller scales. Finally $A_{ij}^{-1}$ is the inverse of matrix $A_{ij}$ given by
\begin{equation}
A_{ij} = \sum_n \frac{(\hat{x}_i \cdot \hat{r}_n)(\hat{x}_j \cdot \hat{r}_n)}{(\sigma_n^2 + \sigma_\star^2)}.
\end{equation}
In practise when calculating the ML weights the first step is to calculate the $A_{ij}$ matrix, taking advantage of the symmetry $A_{ij} = A_{ji}$. The inverted matrix $A_{ij}^{-1}$ is then computed, and the weights $w_{i,n}$ are calculated. This is a fairly simple process, and is cheap in computation time needed.
\subsection{Minimum Variance}
For the Minimum Variance estimator, first an ideal survey is constructed by generating $x$,$y$,$z$ coordinates uniformly randomly in the range $[-4R_I;4R_I]$ and then drawing points according to the distribution $n(r) \propto r^2 \exp{(-r^2/2R_I^2)}$. This constructed ideal survey is spherically symmetric and isotropic. It is constructed such that the window function of the MV method is sensitive in the range where we wish to probe the bulk flow, namely on scales of $R_I$. In order to stay consistent $R_I$ will be set to 50 Mpc $h^{-1}$ in this work, unless otherwise stated. The number of points in the constructed ideal survey is set to 1200 throughout this work. It was found that increasing the number of points in the ideal survey beyond 1200 did not contribute to the stability of the MV method but only served to increase the already considerable computation time.\\

For readability matrix notation is used so that $w_{i,n}$ becomes column matrix $\mathbf{w}_i$ of $n$ elements. $\mathbf{w}_i$ is computed with
\begin{equation}
\mathbf{w}_i = (\mathbf{G} + \lambda \mathbf{P})^{-1} \mathbf{Q}_i.
\end{equation}
$\mathbf{G}$ is a symmetric square $n$ by $m$ matrix where $n$ and $m$ correspond to the $n$'th and $m$'th measurement. The matrix $\mathbf{G}$ is the covariance matrix for the individual velocities $S_n$ and $S_m$. In linear theory we can write the matrix elements $G_{nm}$ as a sum of two terms
\begin{align}
G_{nm} &= \langle S_n S_m \rangle \\
&= \langle v_n v_m \rangle + \delta_{nm}(\sigma_\star^2  + \sigma_n^2).
\end{align}
The second term is known as the noise term and is the Kronecker delta function; 0 for $n \neq m$ but $\sigma_\star^2 + \sigma_n^2$ when $n=m$. The first term is the geometry term which is given by
\begin{equation}
\langle v_n v_m \rangle = \frac{\Omega_m^{1.1}H_0^2}{2\pi^2} \int \mathrm{d}k \, P(k) \, f_{mn}(k)
\end{equation}
where $H_0$ is the Hubble constant in units\footnote{Which is always 100, per definition of $h = (H_0 / 100)$ km$\,$s$^{-1}\,$Mpc$^{-1}$.} of $h$ km s$^{-1}$ Mpc$^{-1}$, and $\Omega_m^{1.1}$ is the growth of structure parameter $f^2 \approx \Omega_m^{1.1}$. $P(k)$ is the matter power spectrum, which in this work is calculated using {\sc copter} \citep{2009PhRvD..80d3531C, 2000ApJ...538..473L, 2012JCAP...04..027H}. The function $f_{mn}(k)$ is the angle averaged window function which is explicitly given as
\begin{equation}
\label{eq:angavwinfuncnasty}
f_{mn}(k) = \int \frac{\mathrm{d}^2 \hat{k}}{4\pi} (\hat{\mathbf{r}}_n \cdot \hat{\mathbf{k}})(\hat{\mathbf{r}}_m \cdot \hat{\mathbf{k}}) \times \mathrm{exp}[ik\hat{\mathbf{k}} \cdot (\hat{\mathbf{r}}_n - \hat{\mathbf{r}}_m)].
\end{equation}
Although Eq. \ref{eq:angavwinfuncnasty} is often quoted in the literature as the function used to calculate $f_{mn}(k)$ it is far from being a practical expression and in reality the expression used is from \cite{2011PhRvD..83j3002M} who showed that we can express the angle averaged window function as
\begin{equation}
\label{eq:fnmk}
f_{mn}(k) = \frac{1}{3} \mathrm{cos}( \alpha(j_0(kA) - 2j_2(kA))) + \frac{1}{A^2} j_2(kA)r_n r_m \mathrm{sin}^2(\alpha)
\end{equation}
where
\begin{equation}
A = (\,r_n^2 + r_m^2 - 2 r_n r_m \mathrm{cos}(\alpha) \, )^{0.5}
\end{equation}
and $\alpha$ is the angle between the $n$'th and $m$'th galaxy given by
\begin{equation}
\alpha = \mathrm{arccos}(\hat{\mathbf{r}}_n \cdot \hat{\mathbf{r}}_m).
\end{equation}
The $j_0(x)$ and $j_2(x)$ functions are spherical Bessel functions given by
\begin{equation}
j_0(x) = \frac{\mathrm{sin}(x)}{x} \,\, , \, \, j_2(x) = \left( \frac{3}{x^2} - 1 \right) \frac{\mathrm{sin}(x)}{x} - \frac{3 \,  \mathrm{cos}(x)}{x^2}.
\end{equation}
Putting all this together gives us the $G_{nm}$ elements. Finding the $P_{nm}$ elements of $\mathbf{P}$ is then fairly simple as it is simply the $k=0$ limit of $f_{nm}$ which is
\begin{equation}
P_{nm} = \frac{1}{3} \mathrm{cos}(\alpha).
\end{equation}
The principal idea of the MV method is to minimise the variance between the bulk flow measured by the galaxy survey and the bulk flow that would be measured by an ideal survey. The $\mathbf{G}$ and $\mathbf{P}$ matrices are the components of the weight that take as input the measured data. The last component, the $\mathbf{Q}$ matrix, takes as input the position and peculiar velocities from the galaxies of the constructed ideal survey. It is calculated in much the same way as the $G_{nm}$ elements with the $Q_{i,n}$ elements being given by
\begin{equation}
Q_{i,n} = \sum_{n'=1}^{N'} w'_{i,n'} \langle v_{n'} v_n \rangle
\end{equation}
and
\begin{equation}
\langle v_{n'} v_n \rangle = \frac{\Omega_m^{1.1} H_0^2}{2\pi^2} \int \mathrm{d}k \, P(k) \, f_{n'n}(k),
\end{equation}
where $f_{n'n}(k)$ is analogous to Eq. \ref{eq:fnmk} but with the difference that $n'$ and $n$ run over the galaxies in the constructed ideal survey, in contrast to $n$ and $m$ that run over the galaxies from the actual observed galaxies of our survey. The ideal weights $w'_{i,n'}$ will be given by
\begin{equation}
w'_{i,n'} = 3 \frac{\hat{\mathbf{x}}_i \cdot \hat{\mathbf{r}}_n}{N_{\rm ideal}}
\end{equation}
where $N_{\rm ideal}$ is the total number of galaxies in the constructed ideal survey.\\

The final step is to solve for the value of $\lambda$, which is a Lagrange multiplier inherent from the minimisation process. It enforces the normalisation constraint
\begin{equation}
\sum_m \sum_n w_{i,n} w_{i,m} P_{nm} = \frac{1}{3}.
\end{equation}
A simple method to solve for $\lambda$ is to vary $\lambda$ and calculate the above sum, until a value for $\lambda$ that makes the above equality true is found.\\

Calculating the MV bulk flow vector is a rather involved process and is orders of magnitude more expensive computationally than the ML estimator. In this work the analysis is done using mainly the ML estimator, with the MV estimator only being tested in a more limited scenario. If computation time was no concern then the full analysis could be carried out for the MV estimator as well.\\

The implementation of the MV estimator used in this work is based on that of Dr. Morag Scrimgeour which is available at https://github.com/mscrim/MVBulkFlow.

\section{Mock Galaxy Surveys versus Dark Matter Halos}
\label{app:mockvsdm}
\begin{figure}
  \centering\includegraphics[width=\linewidth,height=2.5in]{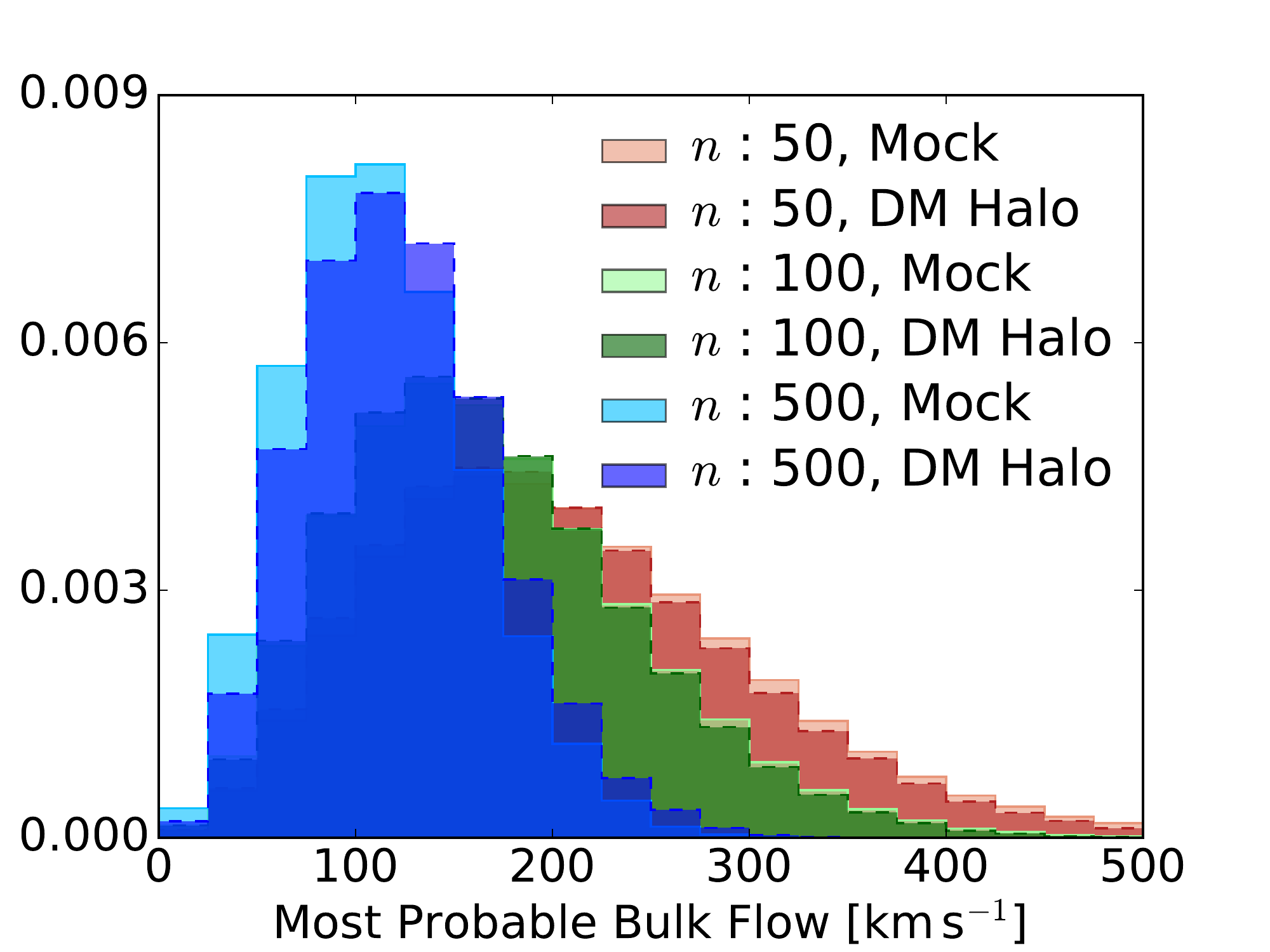}
  \caption{ML bulk flow magnitude distributions for SDSS-III mock galaxy catalogue subsamples and DM halo subsamples, both taken from the same position in the full HR2 simulation. The bulk flow magnitude distributions for the DM halo subsamples are labelled `DM Halo', with the distributions for the SDSS-III mock catalogue samples labelled `Mock'. The individual pairs of bulk flow magnitude distributions (e.g. $n=500$, $n=100$ and $n=50$) all show similar behaviour in their bulk flow velocity distributions.}
\label{fig:compvarcomp}
\end{figure}

As explained in section \ref{sec:hori2} the full HR2 dataset consists of DM halos, not individual galaxies. To test that this does not affect our results, we apply a mock SDSS-III galaxy catalogue produced from the HR2 cosmological DM halo simulation. This mock catalogue lies in a sphere with radius 1 Gpc $h^{-1}$ and origin at ($x,y,z$) = (1.8, 1.8, 1.8) Gpc $h^{-1}$. From the full HR2 DM halo simulation we slice a sphere that also has radius 1 Gpc $h^{-1}$ and origin at ($x,y,z$) = (1.8, 1.8, 1.8) Gpc $h^{-1}$. The distributions of bulk flow magnitudes using the ML estimator are then calculated for both the SDSS-III mock catalogue and the sliced sphere of DM halos. The distributions are shown in Figure \ref{fig:compvarcomp} and the most probable and RMS values are shown in Table \ref{tab:sampcompvarcomp}. We see  that for the same number of galaxies per bulk flow, $n$, the distributions look very similar. From Figure \ref{fig:compvarcomp} and Table \ref{tab:sampcompvarcomp} we can see that the distributions of bulk flow magnitudes, as well as their most probable values and RMS values, are in good agreement. This shows that it is indeed possible to use the DM halos of the full HR2 simulation to perform our analysis, including investigating the effects of survey geometry on the measurements of bulk flow magnitudes.
\begin{table}
\centering
\begin{tabular}{| c | c | l |}
\hline $ $ & SDSSIII Mock & DM Halo \\
  \hline \shortstack{} &
   \shortstack{$n:50$ -  (180$^{+99}_{-83}$)km s$^{-1}$\\ $n:100$ - (147$^{+79}_{-66}$)km s$^{-1}$\\ $n:500$ -  (101$^{+52}_{-44}$) km s$^{-1}$} &
  \shortstack{(180$^{+99}_{-83}$)km s$^{-1}$\\(145$^{+77}_{-65}$)km s$^{-1}$\\(110$^{+55}_{-47}$)km s$^{-1}$}  \\
  \hline
\end{tabular}
\caption{Most probable bulk flow with upper and lower 1-$\sigma$ bounds for bulk flow magnitude distributions of SDSSIII mock survey galaxy catalogue and DM halo slice of the full HR2 simulation, for varying number of galaxies per bulk flow calculation, $n$. The numbers should be compared across horizontally. All the numbers are within 0.1 $\sigma$ of each other, which shows that using DM Halos gives comparable results to using a mock galaxy catalogue.}
\label{tab:sampcompvarcomp}
\end{table}

\section{Estimating Peculiar Velocity Measurement Uncertainty}
\label{app:pecveluncer}
To estimate the peculiar velocity measurement uncertainty, $\sigma_{v,Ia}$, as a function of redshift we follow the approach of \cite{2011ApJ...741...67D}. Using the terminology of \cite{2011ApJ...741...67D} the measurement uncertainty is
\begin{equation}
\sigma_{v,Ia} = c  \cdot \sigma_z  = c \cdot \sigma_\mu \cdot \frac{\ln{(10)}}{5} \frac{\bar{z}(1 + \bar{z}/2)}{1 + \bar{z}}
\end{equation}
where $c$ is the speed of light in vacuum, $\bar{z}$ is the recession redshift and $\sigma_\mu$ is the uncertainty on the distance modulus measurement. To obtain an estimate for the peculiar velocity measurement uncertainty one has to assume a value for $\sigma_\mu$, we have chosen to set $\sigma_\mu=0.1$ throughout this paper, as it is the optimistic value of $\sigma_\mu$ that modern type Ia SNe surveys can achieve, although it is a bit lower than what was possible for legacy surveys where a value of $\sigma_\mu = 0.15$ would be more appropriate. To reiterate the point made in section~\ref{sec:discussion}, using a larger uncertainty in the peculiar velocity measurements will only increase the variance in each component of the bulk flow vector, and any potential biases. Hence by adopting an optimistic error, we are in fact being conservative in our estimates of potential biases.
\end{document}